\documentclass[10pt,letterpaper]{article}
\usepackage{array}
\usepackage{amsmath}
\usepackage{graphicx}
\begin{document}

%%%%%%%%%%%%%%%%%% title page information %%%%%%%%%%%%%%%%%%
\title{Two-port multimode interference reflectors based on aluminium mirrors in a thick SOI platform}

\author{J. S. Fandi\~no$^{1,*}$, J. D. Dom\'enech$^2$ and P. Mu\~noz$^{1,2}$ \\
$^1$ Optical and Quantum Communications Group, \\
Universitat Polit\`ecnica de Val\`encia, 46022 Val\`encia, Spain.\\
$^2$VLC Photonics S.L., Camino de Vera S/N, 46022 Val\`encia, Spain.\\
$^*$jasanfan@iteam.upv.es}

\date{}

\maketitle

\section{Abstract}
Multimode interference reflectors (MIRs) were recently introduced as a new type of photonic integrated devices for on-chip, broadband light reflection. In the original proposal, different MIRs were demonstrated based on total internal reflection mirrors made of two deep-etched facets. Although simpler to fabricate, this approach imposes certain limits on the shape of the field pattern at the reflecting facets, which in turn restricts the types of MIRs that can be implemented. In this work, we propose and experimentally demonstrate the use of aluminium-based mirrors for the design of 2-port MIRs with variable reflectivity. These mirrors do not impose any restrictions on the incident field, and thus give more flexibility at the design stage. Different devices with reflectivities between~0~and~0.5 were fabricated in a 3~um thick SOI platform, and characterization of multiple dies was performed to extract statistical data about their performance. Our measurements show that, on average, losses both in the aluminium mirror and in the access waveguides reduce the reflectivities to about~79\% of their target value. Moreover, standard deviations lower than $\pm$5\% are obtained over a 20~nm wavelength range (1540-1560~nm). We also provide a theoretical modelling of the aluminium mirror based on the effective index method and Fresnel equations in multilayer thin films, which shows good agreement with FDTD simulations.

\section{Introduction}
Multimode interference couplers (MMIs) are one of the most common of the passive building blocks that comprise many modern photonic integrated circuits (PICs). Compact yet tolerant against fabrication errors \cite{Besse1994}, MMIs are very often the natural choice for the implementation of broadband, integrated optical couplers with variable splitting ratios \cite{Soldano1995,Besse1996,Domenech2014}. Their apparently simple geometry can be deceiving, however, as MMIs are in fact very versatile devices. They can find application as high port-count \cite{Kwong2012}, ultracompact \cite{Ortega-Monux2013} and ultra broadband power splitters \cite{Maese-Novo2013}; low crosstalk crossings \cite{Chen2006}; reconfigurable optical switches \cite{Wang2006}; polarization beam splitters \cite{Hosseini2011}; optical multiplexers/demultiplexers \cite{Hu2011}, phase locking of laser arrays \cite{Fukuda2009} as well as phase diversity elements for coherent optical communications \cite{Kunkel2009} or all-optical OFDM receivers \cite{Takiguchi2012}. Furthermore, it is interesting to note that the spatial Talbot effect behind multimodal interference in planar waveguides shares very striking similarities with other physical phenomena \cite{Berry1996}, such as quantum wave-packet evolution \cite{Kaplan2000} and optical pulse propagation through a medium with first order dispersion \cite{Azana2001}.

More recently, yet another type of passive component based on multimode interference was introduced by Xu~et~al. \cite{Xu2009,Kleijn2013}: the multimode interference reflector (MIR). Unlike conventional MMIs, which as mentioned above are commonly employed as power splitters/combiners, MIRs were conceived to be used as broadband on-chip mirrors. They are thus an interesting alternative to already existing reflective elements, such as DBRs, Sagnac loops and cleaved facets, since they inherit the robustness and broadband behaviour of MMIs, but still can be placed anywhere on the chip. In their original work, Xu~et~al. proposed MIRs that employ a reflective element based on two deep-etched facets at a 45$^{\circ}$ angle, which avoids the use of extra fabrication steps. However, rounding effects due to lithography and etching reduce the reflectivity of the mirror towards the center, where the tip is located, introducing extra losses and a possible distortion of the reflected interference pattern \cite{Xu2009,Kleijn2013}. In order the avoid this undesirable effect, the amount of light intensity near the tip of the mirror should be kept as low as possible, which in turn implies that not every MMI can be adapted to create its corresponding MIR without incurring in a power penalty.

In this work, we report on the design and experimental characterization of 2-port MIRs based on aluminium mirrors in a 3 um thick SOI platform. The mirrors are fabricated in this case by deposition of a $\simeq$150~nm thick aluminium film on top of an intermediate silica layer ($\simeq$250~nm), which provides maximum resonant reflection at 1550 nm. Although it requires a higher number of fabrication steps, our approach has the advantage that no limitation exists on the interference pattern formed at the mirror surface. As a consequence, more MMI types can be accommodated to form their equivalent MIRs, providing the designer with a higher degree of flexibility. The same metal mirrors have been recently employed for example in the experimental demonstration of MMI resonators, where two access waveguides of an MMI are replaced by reflecting surfaces in order to implement a ring resonator-like response \cite{Cherchi2015}. Several 2-port MIRs fabricated in this platform with target reflectivities between 0 and 0.5 are experimentally demonstrated to proof the feasibility of our approach.

This work is organized as follows. First, the general concept behind a general M-port MIR is reviewed. Then, and because their greater practical importance, 2-port MIRs based on aluminium mirrors are explained in detail. We give general rules about how arbitrary reflectivities between 0 and 1 might be attained in these structures by exploiting the well-known butterfly 2x2 MMIs conceived by Besse~et~al. \cite{Besse1996}. Next, a simple quasi-analytical model for the aluminium mirror is presented, which combines the effective index method and Fresnel equations in multilayered media. Third, the design of the 2-port MIRs is briefly described. Finally, the experimental characterization of the devices is presented, along with a summary of the main results and conclusions.

\section{Operation principle}
The main idea behind MIRs was first described in \cite{Xu2009} for a 1-port device based on an 1x2~MMI, and was later extended by Kleijn~et~al. for an arbitrary number of input ports \cite{Kleijn2013}. Basically, MIRs are formed by adapting an NxM~MMI, where one of its ends is terminated with a suitable reflective structure. This includes aluminium mirrors or two deep-etched facets at a~45$^{\circ}$ angle, for example. If the mirror is perfect and has no losses, all the forward-propagating modes in the multimode section will be coupled to their corresponding backward-propagating ones. As a consequence, the same (mirrored) interference pattern that otherwise would be formed at the output plane of the original MMI will appear reflected at the access plane of the MIR. If the locations of the input waveguides ($x_{i}$) of the MIR match with the positions of these reflected images, then all the reflected power is captured again, leading to an M-port MIR. It is very important to note that the positions of these reflected images critically depend with the type of mirror employed. Let us denote as $\psi^{-}(x,0)$ the reflected interference pattern at the input of an M-port MIR when a mirror is employed, and $\psi^+(x,L)$ the interference pattern that is formed at the output of its equivalent NxM~MMI when no mirror is present. If two deep-etched facets at~45$^{\circ}$ are used, then this pattern appears mirrored with respect to the center of the MIR ($\psi^-(x,0) = \psi^+(-x,L)$), as explained in \cite{Kleijn2013}. However, this does not happen in a metal-based mirror. In this last case, the whole process can be understood as if the original MMI was folded over with respect to an axis located at half of its length~(L/2). That is, $\psi^-(x,0) = \psi^+(x,L)$.
\begin{figure}[t]
\centering\includegraphics[width=\columnwidth]{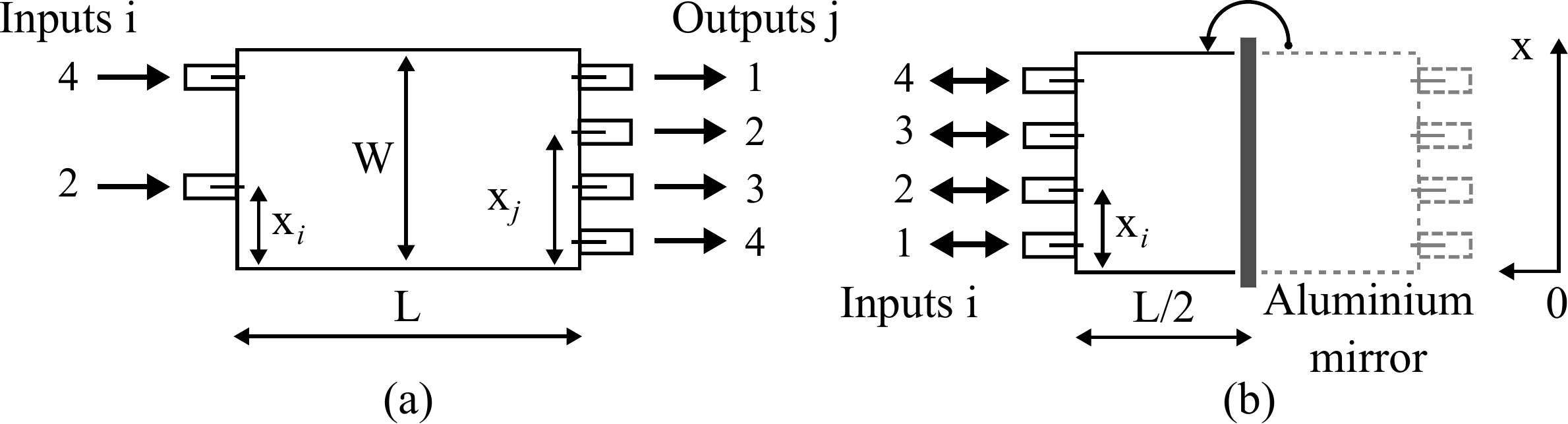}
\caption{Example of how an MIR is formed. (a) Original 2x4 MMI. (b) Equivalent 4-port MIR after introduction of the aluminium mirror.}
\label{Fig1}
\end{figure}

As an example, consider the 2x4~MMI depicted in Fig.~\ref{Fig1}, which is a well-known structure typically used as a~90$^{\circ}$ optical hybrid \cite{Seimetz2006}. As it has already been mentioned, this MMI can be turned into an equivalent MIR with a metal mirror if, for all outputs~(j), there exists an input~(i) such that the position of the input matches that of the output ($x_{i}^{in} = x_{j}^{out}$). In this case, only inputs~2 and~4 match with outputs~1 and~3. Therefore, two extra inputs (1~and~3) need to be added with respect to the original MMI design in order to make sure that all light is extracted from the device, thus preventing unwanted reflections \cite{Pennings1994}. We insist on the fact that, if two deep-etched facets at a~45$^{\circ}$ angle were used, then the positions of the output images would be mirrored with respect to the center of the MIR. In this example, the device is symmetric and all the light would be captured again with the same input positions. However, this is not always the case, and depends on the MMI type that is being adapted.

\section{2-port MIRs with arbitrary reflectivity}
Now that the basics of M-port MIRs have been explained, we will focus our attention in a subfamily of these devices: 2-port MIRs. As explained in \cite{Kleijn2013}, 2-port MIRs can be designed to feature an arbitrary power reflection/transmission ratio~($\rho/\tau$), thus making them useful for a variety of applications in PICs, including on-chip reflectors for integrated lasers as well as arbitrary response AWG-based filters \cite{Gordon2015,Gargallo2014}, to name just a few. In this work, $\rho/\tau$ is defined as the fraction of reflected optical power~($\rho$) divided by the fraction of transmitted optical power~($\tau$), in order to be consistent with the definition of power splitting ratio given in \cite{Besse1996}.

\begin{figure}[t]
\centering\includegraphics[width=0.9\columnwidth]{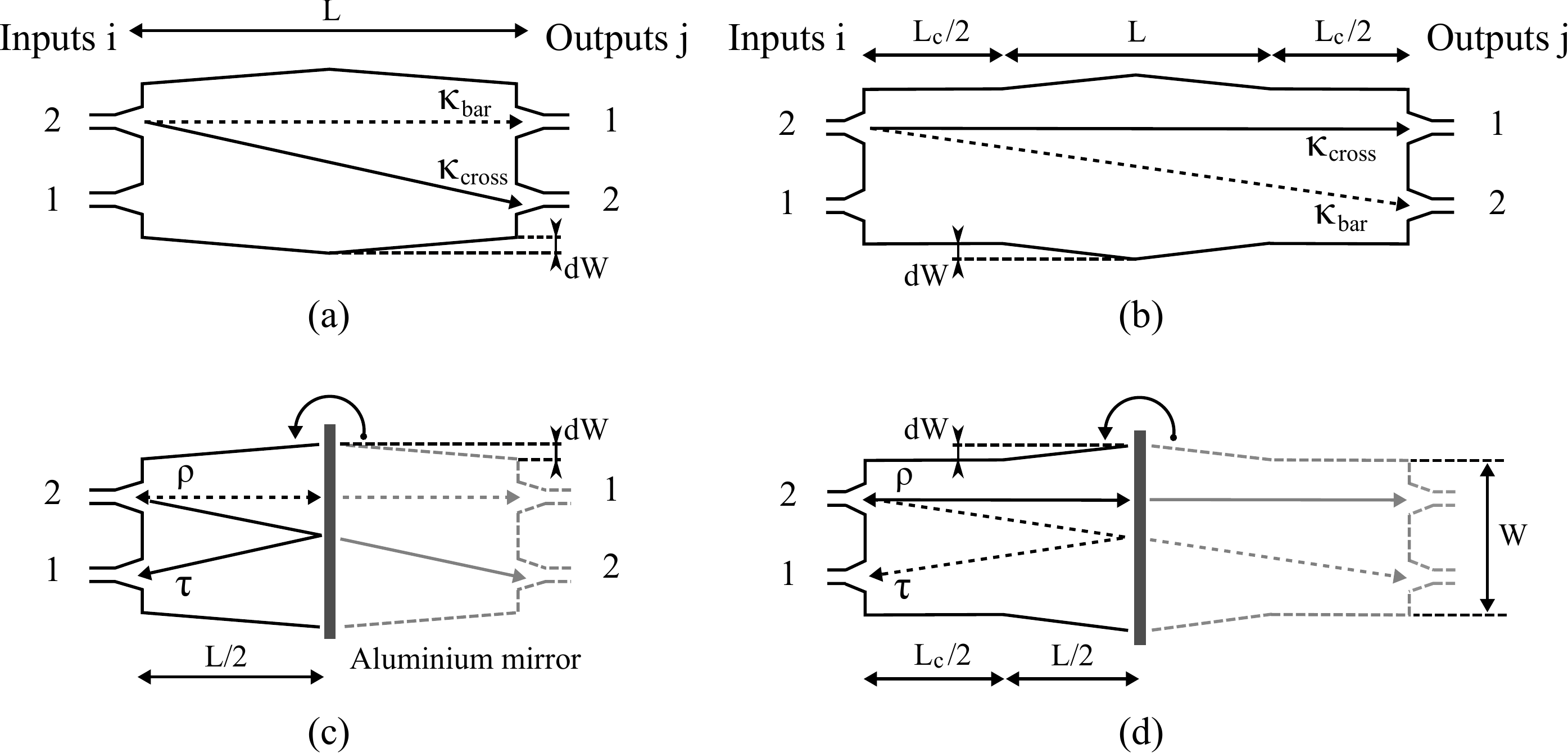}
\caption{(a) and (b) 2x2 butterly MMIs for implementing arbitrary power splitting ratios, without and with crosscoupler, respectively. (c) and (d) Equivalent 2-port MIRs for achieving arbitrary reflectivities.}
\label{Fig2}
\end{figure}

MIRs with arbitrary reflection ratios can be obtained by applying the procedure described in section~2 to the butterfly 2x2~MMIs with arbitrary splitting ratios proposed by Besse~et~al. \cite{Besse1996}. These devices are based on the development by Bachmann~et~al. of the concept of overlapping-image MMIs \cite{Bachmann1995}, which generalizes the restricted interference ideas introduced by Soldano \cite{Soldano1995}. In short, Bachmann~et~al. proved that by carefully positioning the input/output waveguides of an MMI, non-uniform power splitting is possible. For the case of 2x2~MMIs, it was shown that there exist 4 different configurations (namely, types~"A", "B", "C" and "D") that allow for 4 different power splitting ratios (50/50, 100/0, 85/15 and 72/28, respectively). Here, the power splitting ratios are defined as the fraction of power in the cross port ($\kappa_{cross}$) divided by the fraction of power in the bar port ($\kappa_{bar}$), both expressed in \%. An schematic diagram of such devices is shown in Fig.~\ref{Fig2}(a)~and~(b). These 4 devices can then serve as the starting point for the design of a power coupler with arbitrary splitting ratio. This is achieved by adiabatically tapering the width of the MMI, which introduces a mode dependent phase shift that ultimately leads to a change in the intensity of the output self-images. As is also discussed in~\cite{Besse1996}, tapering of types A-D only provides splitting ratios between 50/50 and 100/0. In order to cover the range between~0/100 and~50/50, crosscouplers can be added to exchange the positions of the output ports (see Fig.~\ref{Fig2}(b)). Note that the total length of the crosscoupler ($L_c$) has been split into two sections of length $L_c/2$ located at the input and output of the butterfly MMI, as will be explained shortly.
\begin{table}[t]
\label{Table1}
\begin{center}
\begin{tabular}[t]{>{\centering\arraybackslash}m{0.15\textwidth}ccccc}
\hline
\centering\arraybackslash Schematic & Type & Inputs & L/2 & L$_{\textrm{c}}$/2 & $\rho$/$\tau$ \\ 
\hline 
\includegraphics[width=0.1\textwidth]{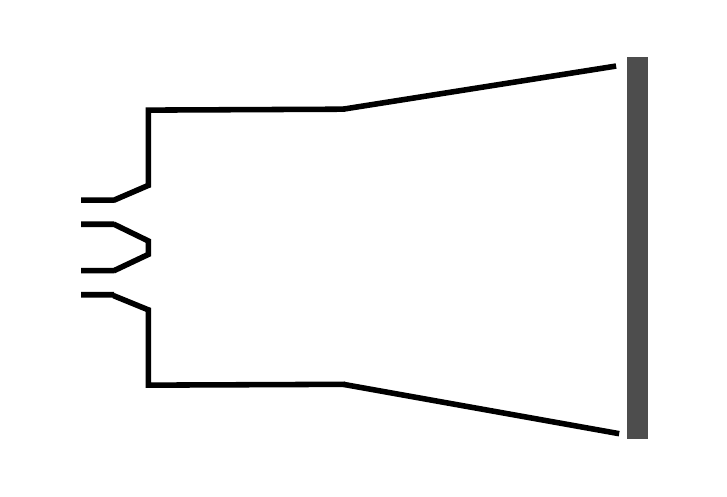} & A & +W/6, -W/6 & L$_{\pi}$/4 & 0 $\mid$ \emph{L$_{\pi}$/2} & 50/50 $\mid$ \emph{50/50} \\ 
 
\includegraphics[width=0.1\textwidth]{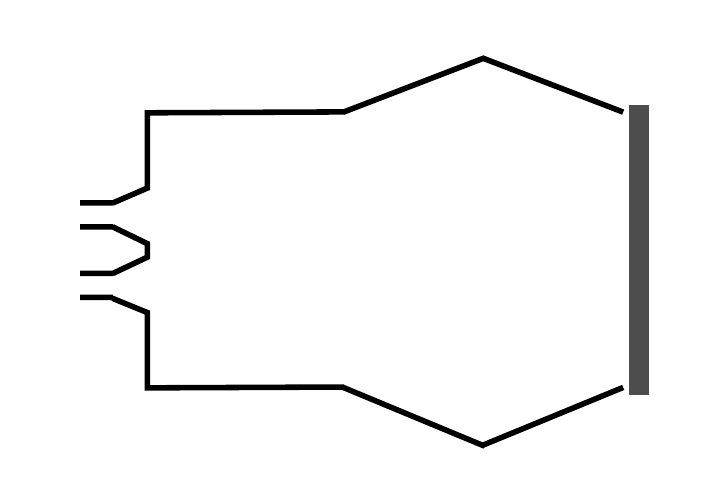} & B & +W/6, -W/6 & L$_{\pi}$/2 & 0 $\mid$ \emph{L$_{\pi}$/2} & 0/100 $\mid$ \emph{100/0} \\ 
 
\includegraphics[width=0.1\textwidth]{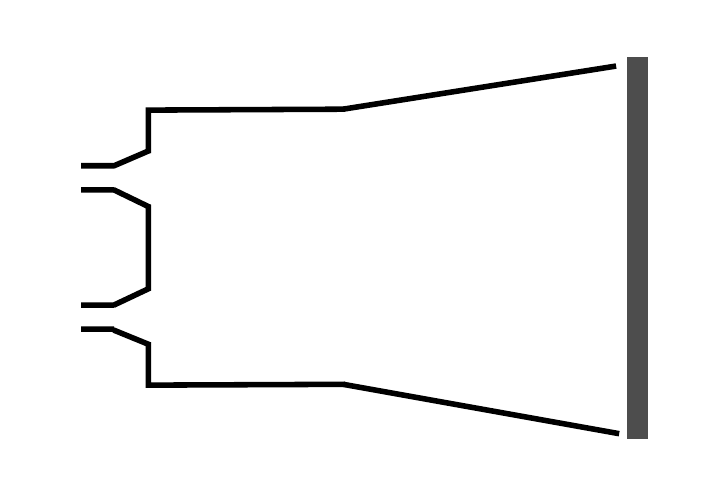} & C & +W/4, -W/4 & 3L$_{\pi}$/8 & 0 $\mid$ \emph{3L$_{\pi}$/2} & 15/85 $\mid$ \emph{85/15} \\ 
 
\includegraphics[width=0.1\textwidth]{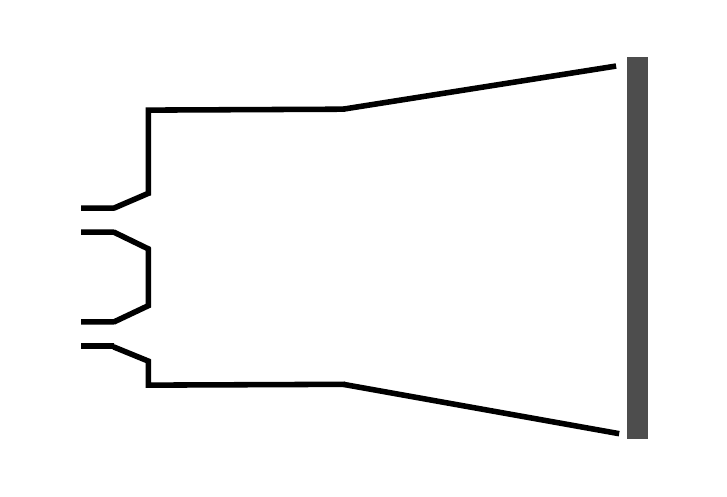} & D & +W/10, -3W/10 & 3L$_{\pi}$/10 & \emph{3L$_{\pi}$/2} & \emph{72/28} \\ 
\hline 
\end{tabular} 
\caption{2-port~MIRs with arbitrary reflectivities. Note that if a crosscoupler is added, then the reflection/transmission ratios change for types~A to~C. The crosscoupler lengths and their corresponding ratios are separated by a vertical bar~($\mid$), where those with a crosscoupler have been highlighted in italics. Type D only works after the introduction of a crosscoupler, as explained in the text. $L_{\pi}$ is defined as~$\lambda_{0}/(2 (\bar{n_{0}}-\bar{n_{1}}))$~\cite{Soldano1995}, where~$\bar{n_{0}}$ and~$\bar{n_{1}}$ are the effective refractive indices of the fundamental and first order modes of the multimode section, respectively. Input positions are measured with respect to the center of the MIR.}
\end{center}
\end{table}

Butterfly 2x2~MMIs can thus be adapted to form 2-port MIRs with arbitrary power reflection/transmission ratios after introduction of the mirror. However, some extra considerations must be taken into account. First, and because of the folding symmetry introduced by the aluminimum mirror, a 2x2~MMI with a power splitting ratio given by $\kappa_{cross}/\kappa_{bar}$ will translate into a 2-port~MIR with a reflection/transmission ratio of $\rho/\tau=\kappa_{bar}/\kappa_{cross}$, as seen Fig.~\ref{Fig2}(c). This is opposite to what happens when using deep-etched facets at a~45$^{\circ}$ angle, since this type of mirror already introduces a reflection of the self-images with respect to the center of the MMI. Second, it must be noted that type "D" devices~(72/28) without crosscouplers do not satisfy the requirements for the input positions explained in section~2. Inputs~1~and~2, located at~-3W/10 and~W/10 with respect to the center, respectively, lead to self-images located at~3W/10 and~-W/10, whose power can not be properly collected. Thus, it is obvious that this device can not be adapted to create an equivalent 2-port MIR. Last, if crosscouplers are added to the basic~A-D types in order to extend the achievable splitting ratios, care must be taken to ensure that the MMI satisfies the required folding symmetry. This can be easily done, however, by noting that the extra length of the crosscoupler ($L_{c}$) can be split into two sections of length~$L_{c}/2$, placed before and after the 2x2~MMI (see Fig.~\ref{Fig2}(b)). The symmetrized structure has exactly the same properties as the original one, as far as the tapering is smooth enough so that the mode conversions remain adiabatic. In that case, each section of the MMI introduces a relative amount of phase shift among the propagating modes, with this effect being independent of the relative order of each section. Adding a crosscoupler thus leads to a mirroring of the output images with respect to the center due to the properties of the self-imaging effect \cite{Soldano1995}. Since types~A to C have inputs which are symmetric with respect to the center of the MIR, this simply exhanges the power splitting values. However, in type~D the crosscoupler leads to output images which are now located at~-3W/10 and~W/10, matching the position of the inputs. As a consequence, the requirements of section~2 are met and it can be used as a 2-port~MIR with a reflection/transmission ratio of~72/28. 

As a summary, Table~1 contains the main geometrical parameters of the 2-port~MIRs described in this section, corresponding to the~4 different types of rectangular MMIs plus their optional crosscouplers. Reflection ratios around those nominal values might be achieved by tapering the devices as described in \cite{Besse1996}, while observing the aforementioned considerations.

\section{Aluminimum mirror modelling}
\begin{figure}[t]
\centering\includegraphics[width=\columnwidth]{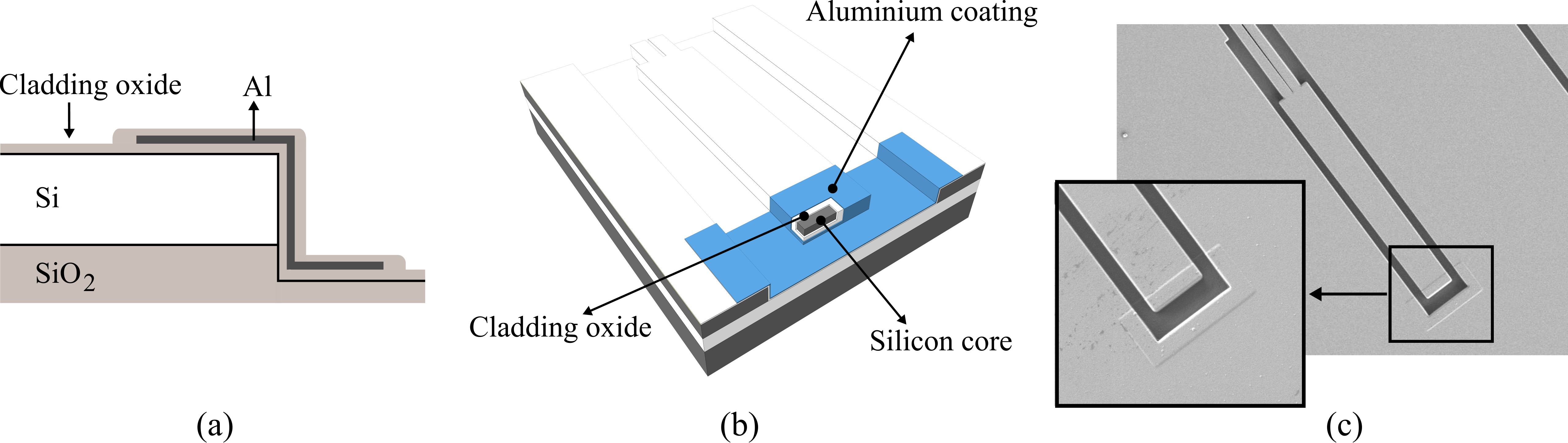}
\caption{(a) Cross-section of the aluminium mirror layer stack. (b) 3D Model of a 2-port MIR with aluminium mirror. (c) SEM picture of a fabricated device.}
\label{Fig3}
\end{figure}
\begin{figure}[t]
\centering\includegraphics[width=0.8\columnwidth]{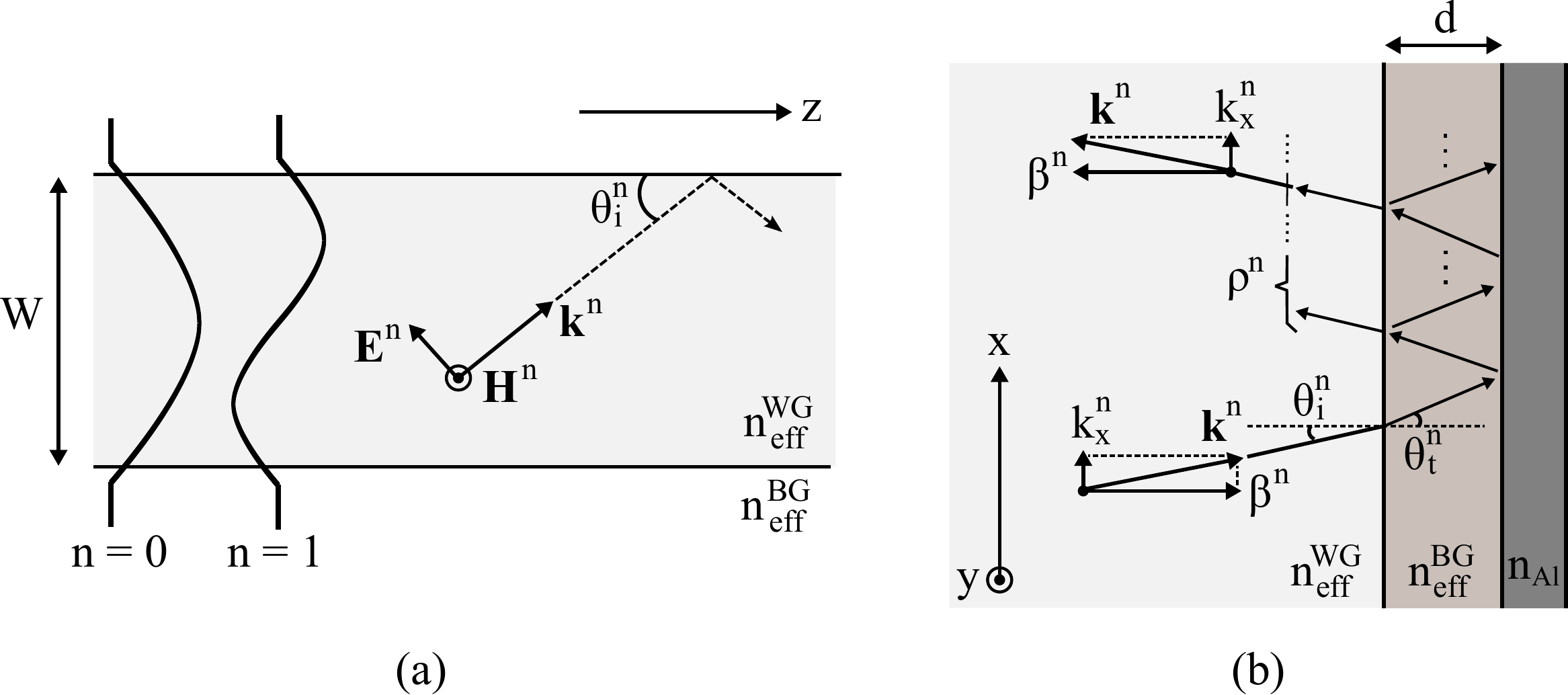}
\caption{(a) Equivalent infinite slab waveguide after reduction of the original MIR cross-section with the effective index method. Each mode (n = 0, 1$\ldots$) is represented here as a plane wave bouncing back and forth at the waveguide interfaces with a certain angle ($\theta_{i}^{n}$). (b) Schematic representation of the resonant reflection of a plane wave (mode) upon incidence on the multilayer structure that comprises the aluminium mirror.}
\label{Fig4}
\end{figure}

A cross-section of the aluminium mirror layer stack that has been employed in this work is shown in~Fig.~\ref{Fig3}(a). Fig.~\ref{Fig3}(b)~and~(c) show a schematic 3D~representation of a 2-port~MIR, as well as an SEM picture of a fabricated device, respectively. In short, the reflectivity of the aluminium coating is enhanced through the use of an intermediate silica layer. Thin-film interference leads to a resonant reflection that is higher than what would be otherwise obtained with aluminium directly deposited on silicon, assuming perpendicular plane wave incidence. More details about this structure can be found in \cite{Cherchi2015}. Since the mirror does not perfectly reflect back all the incident light, it is thus important to model its reflective properties in order to better understand its overall impact on MIR performance. We start our analysis with a commonly employed technique for the simulation and design of MMIs, where the 3D problem is reduced to a simpler 2D geometry with use of the effective index method \cite{Soldano1995}. In this scheme, the vertical dimension is eliminated by finding two effective refractive indices for both the core ($n_{eff}^{WG}$) and the cladding ($n_{eff}^{BG}$). Thus, any mode propagating inside the real (3D) MMI section relates to a mode of the corresponding infinite slab waveguide, whose thickness is equal to the MMI width. According to \cite{Saleh2007}, the modes of a slab waveguide might be understood as plane waves bouncing back and forth on the core/cladding interface (see Fig.~\ref{Fig4}(a)). Each plane wave propagates at a different angle $\theta_{i}^{n}$, which is related to effective index of the \emph{n}th-order mode ($n_{eff}^{n}$) by: 
\begin{equation}
\theta_{i}^{n} = \arccos{ \left( \frac{\beta^{n}}{|\mathbf{k}^{n}|}\right)} = \arccos{ \left( \frac{n_{eff}^{n}}{n_{eff}^{WG}}\right)}
\label{Eq1}
\end{equation}

Under the previous assumptions, the aluminium mirror can be then modelled to first order as an infinite planar interface between three different media: The effective core of the waveguide, the effective cladding and the aluminium layer, as depicted in Fig.~\ref{Fig4}(b). The electric field reflection coefficient in both modulus and phase~($\rho^{n}$) that each mode experiences upon reaching the reflecting surface is now easily obtained by applying the well-known Fresnel equations for multi-layer structures~\cite{Orfanidis2008}. Considering three layers and only TE propagation, which correspond to p-polarized plane waves, the electric field reflection coefficient for the \emph{n}th mode can be expressed as
\begin{equation}
\rho^{n} = \frac{\rho_{12}^{n} + \rho_{23}^{n} \exp{(-j 4 \pi (d/\lambda_{0})n_{eff}^{BG}\cos{\theta_{t}^{n}})}}{1+\rho_{12}^{n}\rho_{23}^{n}\exp{(-j 4 \pi (d/\lambda_{0})n_{eff}^{BG}\cos{\theta_{t}^{n}})}}
\label{Eq2}
\end{equation}
where $d$ is the thickness of the cladding oxide layer, $\lambda_{0}$ is the vacuum wavelength, $n_{eff}^{BG}$ is the effective refractive index of the cladding and $\theta_{t}^{n}$ is the transmitted angle after the first planar interface, which is obtained by applying Snell's law
\begin{equation}
\theta_{t}^{n} = \arcsin{ \left( \frac{n_{eff}^{WG}\sin{\theta_{i}^{n}}}{n_{eff}^{BG}}\right)}
\label{Eq3}
\end{equation}

Finally, $\rho_{12}$ and $\rho_{23}$ relate to the electric field reflection coefficients at the two interfaces, and are given by
\begin{equation}
\rho_{12}^{n} = \frac{(n_{eff}^{BG})^{2} n_{eff}^{WG}\cos{\theta_{i}^{n}} - (n_{eff}^{WG})^{2} n_{eff}^{BG}\cos{\theta_{t}^{n}}}{(n_{eff}^{BG})^{2} n_{eff}^{WG}\cos{\theta_{i}^{n}} + (n_{eff}^{WG})^{2} n_{eff}^{BG}\cos{\theta_{t}^{n}}}
\label{Eq4}
\end{equation}
\begin{equation}
\rho_{23}^{n} = \frac{(n_{Al})^{2} n_{eff}^{BG}\cos{\theta_{t}^{n}} - (n_{eff}^{BG})^{2} [(n_{Al})^{2} - (n_{eff}^{BG}\cos{\theta_{t}^{n}})^{2}]^{1/2}}{(n_{Al})^{2} n_{eff}^{BG}\cos{\theta_{t}^{n}} + (n_{eff}^{BG})^{2} [(n_{Al})^{2} - (n_{eff}^{BG}\cos{\theta_{t}^{n}})^{2}]^{1/2}} 
\label{Eq5}
\end{equation}
where $n_{Al}$ is the (complex) refractive index of aluminium and  $n_{eff}^{WG}$ corresponds to the effective refractive index of the waveguide core. 

In these formulas, it has been assumed by convention that lossy materials exhibit refractive indices with negative imaginary parts. As a consequence, the square root formula of equation~\ref{Eq5} yields two different solutions, of which only the one with negative imaginary coefficient must be considered. Once the electric field reflection coefficients are found, the interference pattern at the input waveguides can be easily found by expanding the input field in terms of the propagating modes of the slab waveguide, as explained in \cite{Soldano1995}, and then multiplying each mode by its corresponding reflection coefficient ($\rho^{n}$) and propagation constant ($\exp{(-j2\pi n_{eff}^{n}L/\lambda_{0})}$). Since the modes propagate at a slightly different angle, it is expected that the mirror will change their relative amplitude and phases, thus affecting the beating pattern at the output.

We performed some numerical simulations to verify our quasi-analytical analysis. The simulation procedure can be described as follows. First, well-known formulas for the refractive indices of silicon and SiO2 were employed to compute $n_{eff}^{WG}$ and $n_{eff}^{BG}$ using a 2D mode solver based on the Film Mode Matching (FMM) method, where the cladding index ($n_{eff}^{BG}$) was optimized using the procedure explained in \cite{Domenech2014}. Then, these indices were fed into an in-house~Matlab\textsuperscript{TM} code that computes both the mode field profiles and propagation constants of an infinite slab with a given width, and then performs the field propagation along the MIR. This includes the mode overlap integrals at both input and outputs, as well as the mode-dependent electric field reflection coefficients given by equation~\ref{Eq2}, where d~$\simeq$~250~nm for this particular platform. Two different MIR types were considered: A type~A device with $\rho/\tau = 50/50$ and a type~B device with $\rho/\tau = 0/100$, whose geometrical parameters can be found in Table~2. Finally, and to check the validity of our model, simulations were compared against high-resolution 2D~FDTD calculations performed with Meep in a~7~node computing cluster~\cite{Oskooi2010}.
\begin{figure}[t]
\centering\includegraphics[width=0.9\columnwidth]{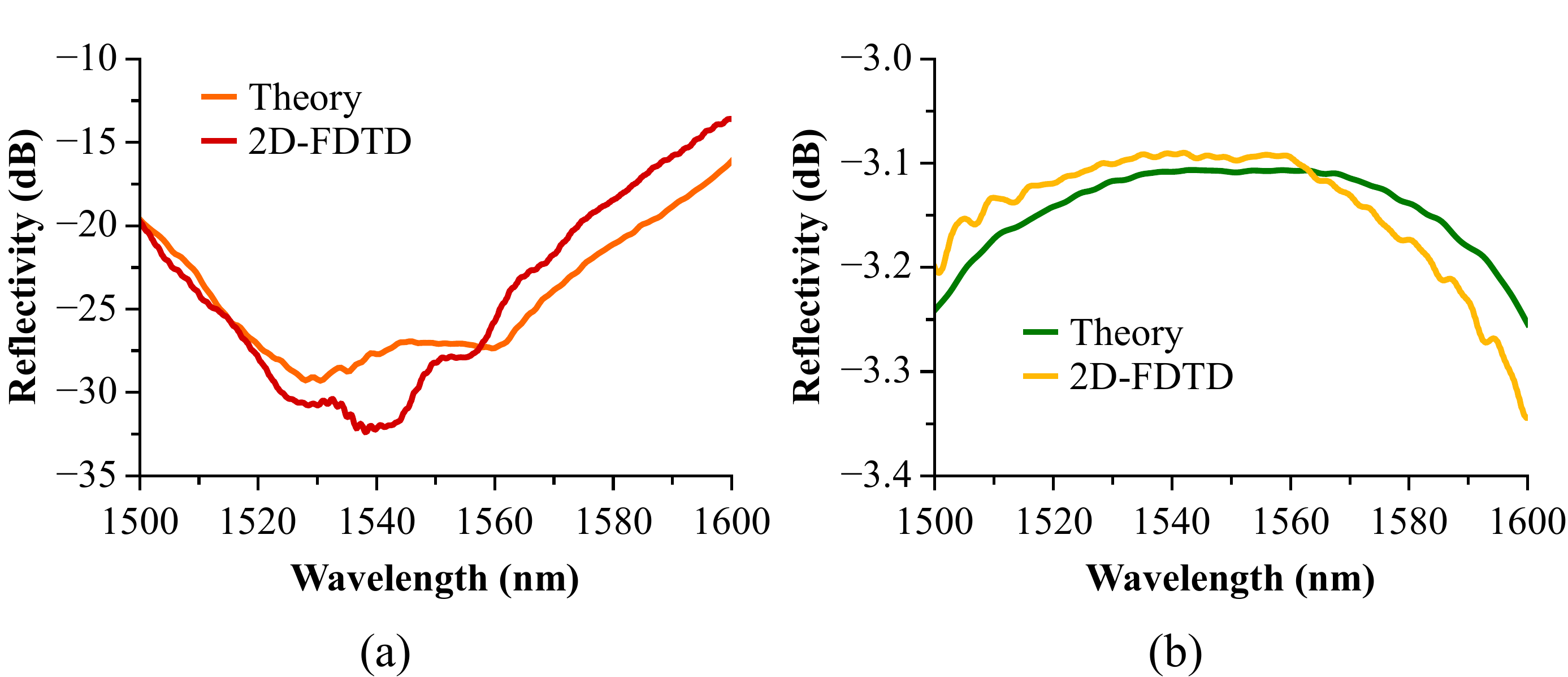}
\caption{(a) Simulated reflectivities for the type~B device of Table~2 ($\rho/\tau=0/100$), both using FDTD (red solid line) and the quasi-analytical theory (orange solid line). (b) Idem for the type~A device of Table~2 ($\rho/\tau=50/50$).}
\label{Fig5}
\end{figure}

Simulation results are shown in Fig.~\ref{Fig5}(a)~and~(b) for type~B and type~A devices, respectively. As it can be seen, our quasi-analytical approach yields very similar results to the more rigorous FDTD method. For type~B, both calculations predict a residual reflectivity lower than -25~dB over more than a 40~nm bandwidth (1520-1560~nm), with the quasi-analytical approach yielding slightly more residual reflected power near the design wavelength (1550~nm). For type~A, a reflectivity higher than -3.2~dB over a 60~nm bandwidth (1520-1580~nm) is achieved in both cases, with about 0.1~dB insertion losses at 1550~nm. These correspond to an effective peak reflectivity of~97.7\%, in good agreement with previously reported simulation results \cite{Cherchi2015}, and points to a negligible performance degradation introduced by the mirror. In practice, however, fabrication imperfections such as surface roughness and changes in the sidewall angle will affect the reflecting performance of the designed mirror. This will not only introduce extra losses, but also a distortion of the reflected field. The distortion will change the relative amplitudes and phases of the backward-propagating modes, which will cause in turn extra losses and imbalance in the self-images formed at the input of the MIR. Thus, designers might need to consider these effects beforehand, and compensate for a possible reduction of the reflectivity during the design stage.

\section{Device design}
\begin{table}[t]
\begin{center}
\begin{tabular}[t]{>{\centering\arraybackslash}m{0.1\textwidth}ccccccc}
\hline
\centering\arraybackslash Schematic & Type & Inputs ($\mu$m) & dW ($\mu$m) & L/2 ($\mu$m) & L$_{\textrm{c}}$/2 ($\mu$m) & Target $\rho$/$\tau$ & Sim. $\rho$/$\tau$ \\ 
\hline 
\includegraphics[width=0.1\textwidth]{./Type_A.pdf} & A & $\pm$1.95 & 0 & 101.92 & 0 & 50/50 & 49.1/49.6\\ 
 
\includegraphics[width=0.1\textwidth]{./Type_A.pdf} & A & $\pm$1.95 & 0.75 & 88.29 & 0 & 40/60 & 39.1/59.1 \\ 
 
\includegraphics[width=0.1\textwidth]{./Type_C.pdf} & C & $\pm$2.93 & -2.785 & 225.77  & 0 & 30/70 & 30.0/68.1\\ 
 
\includegraphics[width=0.1\textwidth]{./Type_C.pdf} & C & $\pm$2.93 & -1.057 & 180.39 & 0 & 20/80 & 19.8/79.0 \\ 
 
\includegraphics[width=0.1\textwidth]{./Type_B.pdf} & B & $\pm$1.95 & 0 & 203.84 & 0 & 0/100 & 0.0/98.9\\ 
\hline
\end{tabular} 
\caption{Physical dimensions and BPM simulation results of the 5 different fabricated devices. The width is common for all designs (11.7~$\mu$m).}
\end{center}
\label{Table2}
\end{table}
In order to experimentally demonstrate the concept, 5 different 2-port MIRs with reflectivities varying between~0~($\rho/\tau = 0/100$) and~0.5~($\rho/\tau=50/50$) were considered. Given the almost negligible effect introduced by the mirror found in the previous section, and the lack of more realistic experimental data, an ideal metallic mirror was assumed in order to simplify the design process. Under the previous assumption, the effect of the mirror is neglected and the MIRs dimensions can be obtained by simply simulating their equivalent butterfly MMIs using the well-known Beam Propagation Method~(BPM). The design is thus reduced to that explained in \cite{Domenech2014}, which can be briely described as follows. First, 5 different power splitting ratios ($\kappa_{cross}/\kappa_{bar}$) in the range between 50/50 and 100/0 were chosen, namely: 50/50, 60/40, 70/30, 80/20 and 100/0. Butterfly MMIs with these splitting ratios do not require an extra crosscoupler, which results in shorter devices with an enhanced operation bandwidth~\cite{Besse1994}. Each of these ratios was then assigned to be implemented by a given butterfly MMI type (A, B or C), which was done by trying to minimize the difference between the target value and the splitting ratio of the canonical device (50/50, 100/0 and 85/15). Please note that, as explained in section~3, these splitting ratios lead to reflection/transmission ratios between 50/50 and 0/100, this is: 50/50, 40/60, 30/70, 20/80 and 0/100. A common width of 11.7~um for all devices was found to be enough to satisfy the minimum spacing between the 2.8~um wide input waveguides, thus ensuring a proper opening of the access plane during lithography and etching. Next, an optimized effective index method and 2D BPM propagations were employed to iteratively find the optimum geometrical parameters of the butterfly MMIs, namely length (L) and width variation (dW). The equivalent MIR geometries can finally be determined by noting that they are equal to that of the designed butterfly MMIs, except for the length, which is half of it (L/2). Table~2 shows the final parameters of the designed devices, together with their simulated performance. Please note that here negative dW correspond with a widening of the MMI body, and viceversa.

\section{Measurement procedure and experimental results}
\begin{figure}[t]
\centering\includegraphics[width=0.8\columnwidth]{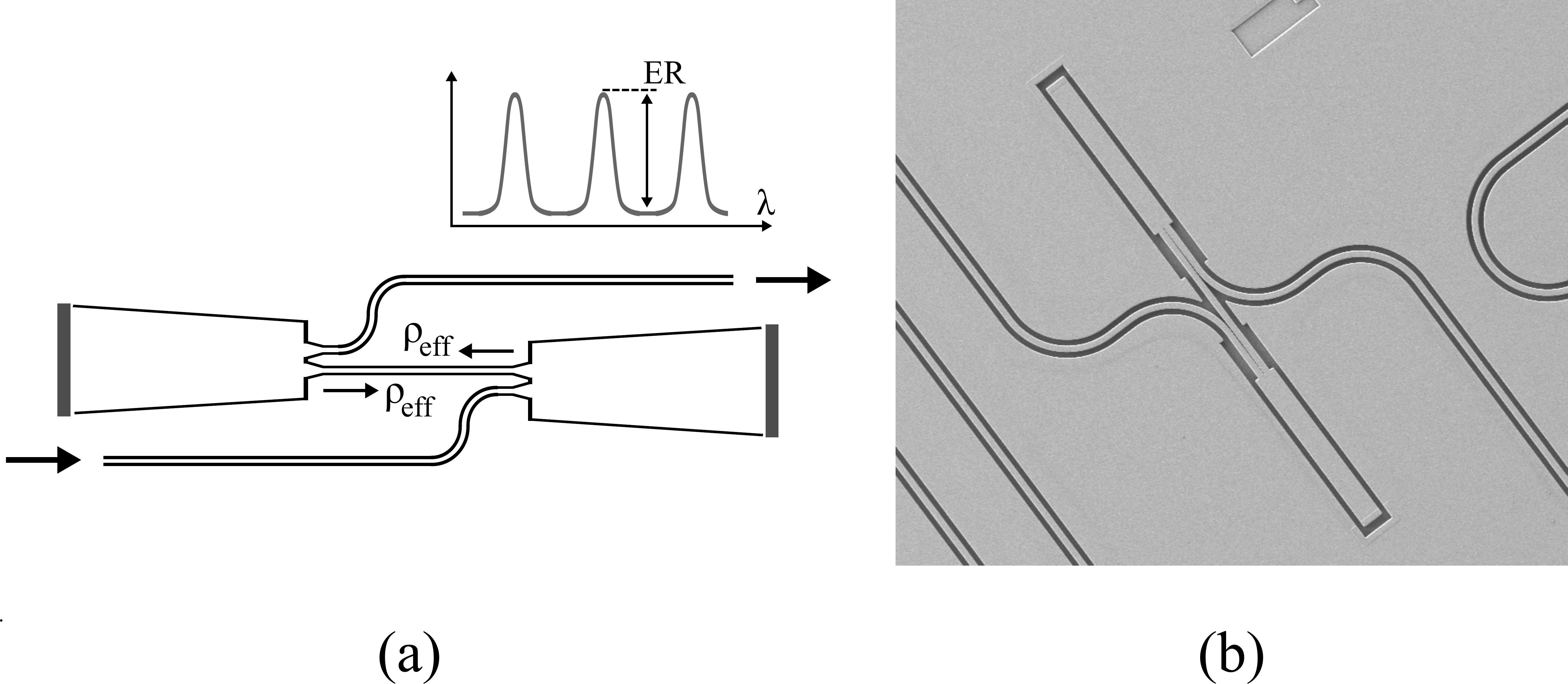}
\caption{(a) Schematic diagram of the structures employed to measure the effective reflectivities of the devices under test. (b) SEM picture of a test structure.}
\label{Fig6}
\end{figure}
The designed devices were fabricated in a commercially available 3~um thick SOI platform. More details about the fabrication procedure and technology capabilities can be found elsewhere in the literature~\cite{Cherchi2015,Solehmainen2006}. Fabry-P\'erot test structures made of a concatenation of two MIRs were employed to extract the wavelength dependent reflectivity, similar to~\cite{Kleijn2013}. Fig~\ref{Fig6}(a) shows an schematic diagram of the test structures, while Fig~\ref{Fig6}(b) is an SEM picture of a fabricated device. The two MIRs form an integrated resonant cavity with a periodic spectrum made of interference fringes. The depth of these fringes, also known as the extinction ratio~(ER), depends on the total effective reflectivity of the MIR under test ($\rho_{eff}$). This includes not only the reflectivity of the metallic mirrors, but also the insertion losses in the access waveguides and the propagation losses within the multimode section. An interesting feature of this indirect approach is that the extinction ratio does not depend on the coupling losses, which can fluctuate from measurement to measurement due to mechanical vibrations and thermal drifts. Since the losses in the connecting straight waveguides can be considered negligible ($\leq$0.1~dB), $\rho_{eff}$ is given by
\begin{equation}
\rho_{eff} = \frac{\sqrt{\text{ER}}-1}{\sqrt{\text{ER}}+1}
\end{equation}
where ER is the measured extinction ratio, in linear units.

The measurement procedure was as follows. A broadband ASE source (ASE-CL-20-S, NP~Photonics) was fiber coupled to a Tholabs'~FiberBench, consisting of a collimator, a free space polarizer and a focusing microscope objective. The free space polarizer was adjusted so that only TE~polarization, parallel to the chip surface, was injected into the chip. Light was collected at the chip output by a lensed fiber and directed into an optical spectrum analyzer (AQ6370C, Yokogawa), where it was recorded and sent to a computer through a GPIB~interface. The transmission spectra of test straight waveguides located nearby on the chip were also collected so as to normalize and compensate for the non-flatness of the source. In total, 9~different chips were measured, with six test structures each (5~MIRs + 1~straight waveguide).
\begin{figure}[t]
\centering\includegraphics[width=\textwidth]{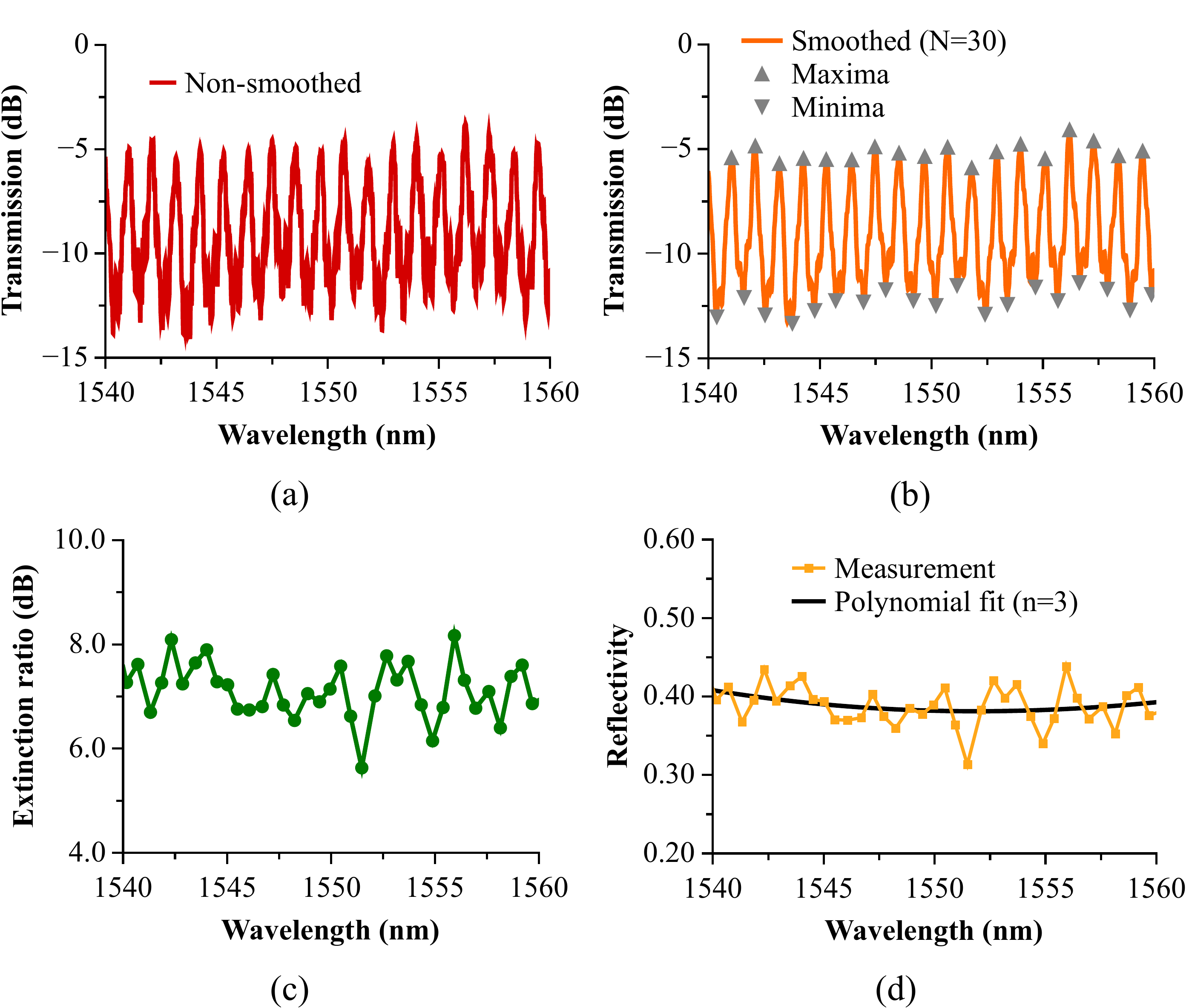}
\caption{(a) Measured spectrum of the type~A device of Table~2 ($\rho/\tau=50/50$), after normalization with a test straight waveguide. (b) Same spectrum, but smoothed with a 30 point moving average ($\simeq$120~pm). Maxima and minima are also shown as grey triangles. (c) Extracted extinction ratios for different wavelengths. (d) Calculated reflectivity and third-order polynomial fit (solid black line).}
\label{Fig7}
\end{figure}

An example of a measured transmission spectrum after normalization is shown in Fig.~\ref{Fig7}(a), corresponding to device~1 in Table~2 (type~A, $\rho/\tau = 50/50$). As it can be seen, the interference fringes are severely affected by spurious ripples. These ripples mainly come from non-negligible reflections between the chip facets and the MIRs. They were in fact also observable in the spectrum of the reference straight waveguides, and are associated with the relatively low performance of the antireflection coating. Unfortunately, they introduce an indetermination in the measured extinction ratio, which in turn causes noisy fluctuations in the spectral dependence of the reflectivity. In order to reduce this unwanted effect, a moving average of 30~points, equivalent to a 120~pm wide spectral window, was applied to the recorded traces. The result of this smoothing can be seen in Fig.~\ref{Fig7}(b), where maxima and minima are also shown. Every two adjacent minima and maxima points can now be used to compute the extinction ratio at an intermediate wavelength, which is plotted in Fig.~\ref{Fig7}(c). Finally, the extinction ratio in linear units is used to compute the effective reflectivity versus wavelength, as shown in Fig.~\ref{Fig7}(d). Even though smoothing significantly reduces the wavelength fluctuations of the reflectivity, it is clear that the ripples are still noticeable. In order to have a better estimate, we exploit the fact that the reflectivity varies slowly with wavelength as predicted by both FDTD simulations and the quasi-analytical model, and fit a third-order polynomial to the measured traces over a 20~nm wavelength range (1540-1560~nm). The result is plotted as solid black line in Fig.~\ref{Fig7}(d). A reflectivity of about 0.4 is obtained over the whole bandwidth, which deviates from the target value of 0.5. This is attributed to the combined effect of both the non-perfect reflectivity of the mirror and the excess losses of the MMI, which are estimated to be around 1~dB in total~($0.4/0.5=0.8$).

\begin{figure}[t]
\centering\includegraphics[width=\textwidth]{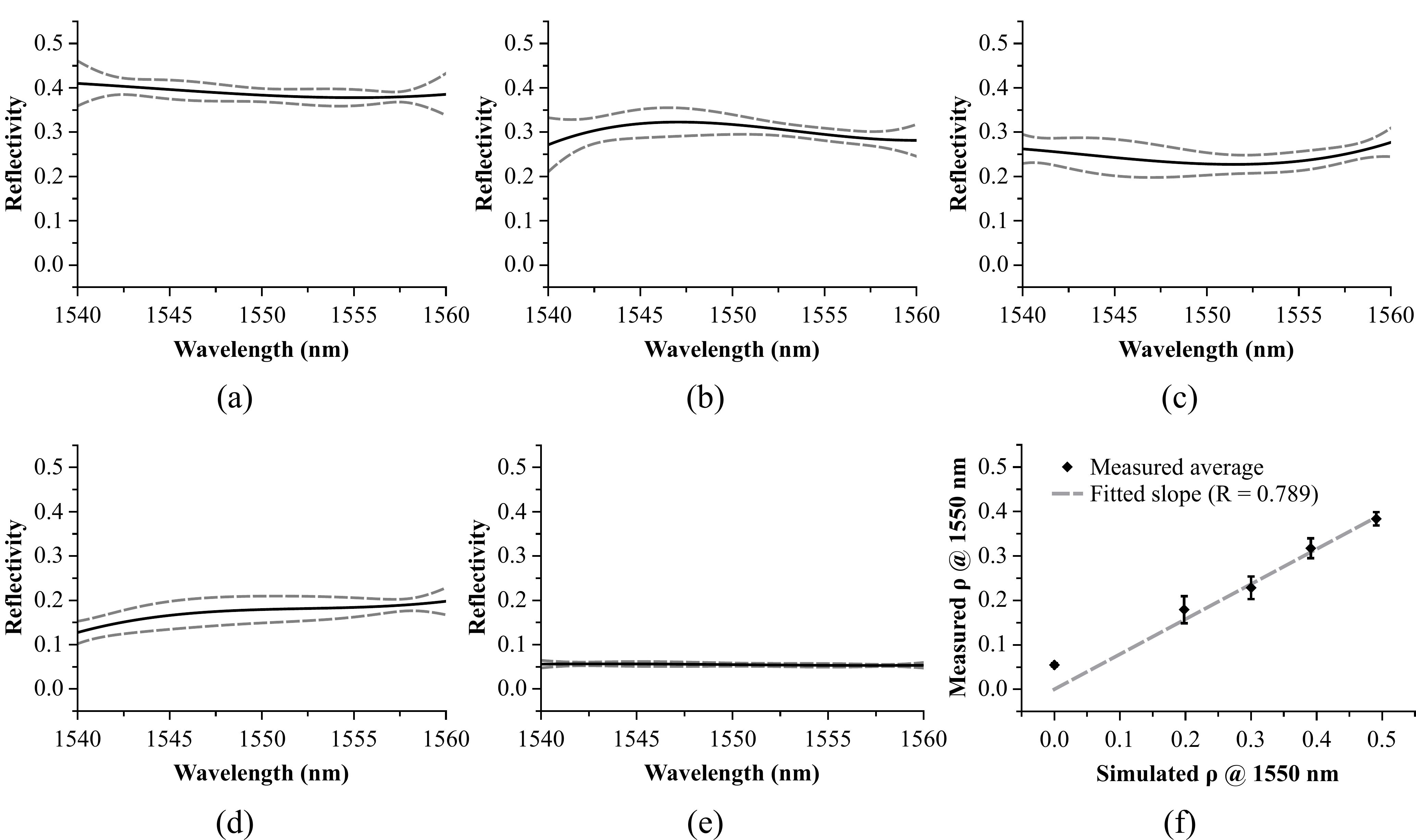}
\caption{(a) to (e) Average reflectivities (solid black lines) and standard deviations (grey dashed lines) for the 5 MIRs shown in Table~2, computed after measuring 9 different dies. (f) Measured average and standard deviations of the reflectivities versus the simulated values at 1550~nm. Grey dashed line: Slope fitted to the data that provides an estimate of the average intrinsic losses both in the mirror and in the access waveguides.}
\label{Fig8}
\end{figure}

Polynomial fittings were performed for all structures in the 9 measured chips, and then these were employed to obtain an statistical estimate of the expected reflectivity fluctuations due to the manufacturing process. Fig~\ref{Fig8}(a)~to~(e) shows the average (solid black line) and standard deviations (grey dashed line) of the wavelength dependent reflectivities for each designed MIR. As it can be seen, all the devices have a maximum standard deviation with respect to the average value of about~$\pm$5\%. The device with a target $\rho/\tau=0/100$ (type~B in Table~2) features an average residual reflectivity of about~5\%, which we mainly attribute to spurious reflections in the chip facets not eliminated during smoothing. This also explains why this device exhibits a negligible standard deviation in the measured reflectivity, as these reflections are not expected to change significantly during manufacturing. Finally, the average reflectivities and standard deviations for all the devices at the nominal operation wavelength (1550~nm) are plotted in Fig.~\ref{Fig8}(f) against the simulated reflectivities of Table~2. By fitting a slope to the data, an estimate of the reflectivity due to the aluminium mirror and the excess losses of the MMI is obtained. As it can be seen in Fig.~\ref{Fig8}(f), the effective reflectivity is around 79\%, similar to other previous works \cite{Xu2009,Kleijn2013,Cherchi2015}. However, we believe this figure could be a bit higher since the applied smoothing procedure results in a systematic, albeit small, reduction of the measured extinction ratio.	

\section{Conclusions}
In this work, multimode interference reflectors (MIRs) based on aluminium mirrors have been designed and experimentally demonstrated in a 3~um SOI platform. Contrary to previous approaches, MIRs based on metal mirrors feature a greater degree of flexibility as they do not have restrictions on the shape of the field pattern formed at the reflecting surface. Several devices with reflectivities varying between 0 and 0.5 were fabricated, and characterization of~9 different dies was performed in order to extract statistical data about their performance. Measurements show average reflectivities to be around~79\% of the target value, mainly due to both non-ideal reflection from the mirror and losses in the access waveguides. Moreover, maximum standard deviations of the reflectivity of about~$\pm5$\% are achieved in a 20~nm (1540-1560~nm) wavelength range for all designs. Finally, we have also provided a thorough theoretical description of these devices. This includes formulas for the design of MIRs with arbitrary reflection/transmission ratios, as well as expressions for the modelling of the aluminium mirror, which are in good agreement with more realistic 2D-FDTD simulations. This type of devices might thus be interesting for the implementation of future fabrication-tolerant, broadband, on-chip reflectors, finding applications in reflective optical filters and integrated lasers, among others.

\section*{Acknowledgements}
This work was supported by projects TEC2010-21337 (ATOMIC), FEDER UPVOV10-3E-492, FEDER UPVOV08-3E-008, TEC2013-42332-P (PIC4ESP), and PROMETEO 2013/012. The work of J. S.Fandi\~no was supported by Grant FPU-2010 (ref: AP2010-1595).

\end{document}